\pdfoutput=1
\documentclass[%
  article,notitlepage,
 reprint,
 amsmath,amssymb,
 aps,
longbibliography,
 floats,
superscriptaddress
]{revtex4-1}

\usepackage{graphicx}
\usepackage{dcolumn}
\usepackage{bm}
\usepackage{color}
\usepackage{amssymb}

\begin{document}

\title{Active rotational dynamics of a self-diffusiophoretic colloidal motor}
\thanks{Electronic Supplementary Information (ESI) available:
  A video showing the active rotational motion of a Janus motor with an asymmetrical catalytic domain (Movie $S1$).}

\author{Shang Yik Reigh}
\email{silee@snu.ac.kr; reigh@hhu.de}
\affiliation{The Research Institute of Basic Sciences, Seoul National University,
Seoul 08826, Republic of Korea}
\affiliation{Institut f{\"u}r Theoretishe Physik II: Weiche Materie, Heinrich-Heine-Universit{\"a}t, 40225 D{\"u}sseldorf, Germany}
\affiliation{Max-Planck-Institut f{\"u}r Intelligente Systeme,
    Heisenbergstra{\ss}e 3, 70569 Stuttgart, Germany}   
\author{Mu-Jie Huang}
\email{mjhuang@chem.utoronto.ca}
\affiliation{Chemical Physics Theory Group, Department of
    Chemistry, University of Toronto, Toronto, Ontario M5S 3H6, Canada}
\author{Hartmut L{\"o}wen}
\affiliation{Institut f{\"u}r Theoretishe Physik II: Weiche Materie,
    Heinrich-Heine-Universit{\"a}t, 40225 D{\"u}sseldorf, Germany}
\author{Eric Lauga}
\affiliation{Department of Applied Mathematics and Theoretical Physics, Center for Mathematical Science, University of Cambridge, Wilberforce Road, Cambridge CB3 0WA, United Kingdom}
\author{Raymond Kapral}
\affiliation{Chemical Physics Theory Group, Department of
    Chemistry, University of Toronto, Toronto, Ontario M5S 3H6, Canada}

\date{\today}

\begin{abstract}
  The dynamics of a spherical chemically-powered synthetic colloidal motor that operates by a self-diffusiophoretic mechanism and
  has a catalytic domain of arbitrary shape is studied using both continuum theory and particle-based simulations.
  The motor executes active rotational motion when self-generated concentration gradients and interactions between the
  chemical species and colloidal motor surface break spherical symmetry. Local variations of chemical reaction
  rates on the motor catalytic surface with catalytic domain sizes and shapes provide such broken symmetry conditions.
  A continuum theoretical description of the active rotational motion is given, along with the
  results of particle-based simulations of the active dynamics. From these results a detailed description of the factors
  responsible for the active rotational dynamics can be given. Since active rotational motion often plays a significant part in the nature of
  the collective dynamics of many-motor systems and can be used to control motor motion in targeted cargo transport, our results should find applications beyond those considered here.
\end{abstract}

\pacs{Valid PACS appear here}

\maketitle

\section{Introduction}
Small self-propelled colloidal particles that use chemical energy derived from their environments to execute directed motion have been the subject of numerous investigations because of their potential applications and the new phenomena that arise in systems of such active colloids~\cite{wangbook:13,SenRev:13,kapral:13,sanchez:14,stark:16,bechinger:16,maass:16}. Colloidal motors with different shapes and sizes that are propelled by various mechanisms have been made and their properties have been characterized (see e.g., Ref.~\cite{paxton04,pope:10,graaf:15,sanchez:16,michelin:17}). The focus of the work described here is on the active orientational motion of motors that operate by phoretic mechanisms~\cite{derjaguin:47,anderson:89}, especially through self-diffusiophoresis~\cite{golestanian:05,kapral:13,colberg14,stark:16,yu:18}.

The self-diffusiophoretic mechanism operates for colloidal particles whose surfaces have catalytic and noncatalytic domains. Under nonequilibrium conditions, chemical reactions on the catalytic portion of the colloid surface generate fuel and product concentration gradients that give rise to pressure gradients on the fluid and shear stresses on the particle derived from the fluid-colloid interaction potentials. Since no external forces are present, momentum conservation leads to fluid flows in the surrounding medium that are responsible for motor motion.

Most studies of diffusiophoretic motors have considered the simplest motor geometry: spherical Janus motors with catalytic and noncatalytic hemispherical caps~\cite{wangbook:13,wang:13,anderson:89,paxton04,wheat:10}. As a consequence of axial symmetry these motors execute active directed motion along the polar axis of the motor but their orientational dynamics is controlled solely by rotational Brownian motion. Janus motors with variable cap sizes that retain the axial symmetry of the colloid have also been studied but, like the Janus motors with hemispherical caps, they cannot execute active rotational motion~\cite{golestanian:07,pope:10,graaf:15,michelin:17}. However, active rotational motion is obtained if the axial symmetry of the spherical colloid is broken by asymmetrical catalytic domains, for example, inhomogeneous catalytic reaction rates~\cite{archer2015} or asymmetrical distribution of catalytic domains on the motor surface~\cite{lisicki:18}.
Active rotational motion has also been studied for Janus motors
with a coupling of electrochemical forces to fluid flow~\cite{holger:19}
or under external field such as gravity~\cite{campbell:17},
nonspherical colloidal motors including L-shaped particles~\cite{bechinger:13,debuyl:19} and dimer aggregates of motors~\cite{ebbens:10,sano:10,snigdha:10,wittmeier:15,majee:17,johnson:17}.

In this paper we investigate self-diffusiophoretic spherical Janus motors with catalytic domains that break the axial symmetry of the colloid and undergo active rotational motion. Experimental realizations of Janus colloids with such catalytic domains have been made by glancing angle metal evaporation~\cite{archer2015}. Catalytic domains made by this process have variable thickness and evidence suggests that propulsion occurs largely through an electrophoretic mechanism where the catalytic activity depends on the thickness of the metal coating~\cite{ebbens:14}, although it has also been suggested that active rotation can  arise solely due to diffusiophoresis. In the following we present the full continuum theoretical description of the self-diffusiophoretic active rotational dynamics of Janus colloids with catalytic domains of arbitrary shape, along with particle-based simulations of the translational and rotational dynamics of such Janus colloids. We show that both the domain shape and spatial variations of the catalytic activity play important roles in determining the nature and magnitude of the active rotational motion.

The outline of the paper is as follows. Section~\ref{sec:cont} gives the general expressions for the linear and angular propulsion velocities, along with the solutions of the reaction-diffusion equations for the concentration fields that enter these formulae. Simulations of the Janus motor dynamics are presented in Sec.~\ref{sec:sims} where the spatial structure of the concentration fields is studied in detail as a function of the catalytic domain shape and spatial dependence of the reaction rate. In addition, various aspects of the active rotational motion and its effects on the motor mean square displacement are presented. Section~\ref{sec:conc} contains the conclusions of the study, while the Appendix provides details of the simulation method.

\section{Continuum theory\label{sec:cont}}
\subsection{Linear and angular motor velocities}
We consider a spherical motor with hydrodynamic radius $R_M$ immersed in fluid with dynamic viscosity $\eta$ and density $\rho$ comprising reactive $A$ (fuel) and $B$ (product) particles. The motor has a catalytic ($C$) region with an arbitrary shape on its surface while the rest of the surface is noncatalytic ($N$). Assuming that catalytic reactions on the motor surface strongly favor the formation of products, we consider only the forward reactions, $C+A \stackrel{k_0}{\rightarrow} C+B$, to occur with an intrinsic reaction rate $k_0$. In order to maintain the system in a nonequilibrium state we also suppose that a chemical reaction $B \stackrel{k_2}{\rightarrow} A$ occurs by a different mechanism in the fluid phase; e.g., by a reaction $E+B \stackrel{k_3}{\rightarrow} F+A$, where the fixed concentration of species $[E]$ is incorporated in the rate constant $k_2=k_3 [E]$ and can be used to vary the value of $k_2$. The introduction of fluid-phase reactions to supply fuel and remove product mimics the way these processes occur in biological contexts, and they can be implemented in {\it in vitro} experiments. They are especially important for the particle-based simulations described below since the system must be maintained in a nonequilibrium state for long time periods to extract reliable statistical data on motor motion~\cite{Huang_etal_2018,Huang_etal_2019}.

The deterministic force and torque on a colloidal motor can be computed from surface averages of the fluid pressure tensor with boundary conditions that account for values of the fluid velocity fields on the surface and diffusiophoretic coupling to the chemical species concentration gradients~\cite{anderson86,anderson:89}. For partial slip boundary conditions on the surface of the colloid, they take the following forms on time scales longer than the hydrodynamic
time, $R^2_M \rho/\eta$:~\cite{Pierre_Kapral_18,Pierre_Kapral_19}
\begin{align}
\bm{F} = \frac{\zeta_t^\circ}{4\pi R_M^2(1+3b/R_M)} \sum_{h=C}^N \lambda_{h} \:\int_S dS \: H_h(\theta,\phi)  \nonumber \\
 \hspace{2pt} \times \Big[ (\mathbf{I} - \hat{\bm{r}}\hat{\bm{r}}) \cdot {\bm{\nabla}} c_B(\bm{r})\Big], \label{eq:F}
\end{align}
\begin{align}
\bm{T} = \frac{3\zeta_r^\circ}{8\pi R_M^3(1+3b/R_M)} \sum_{h=C}^N \lambda_{h} \:\int_S dS \: H_h(\theta,\phi) \nonumber \\
 \times \Big[ \hat{\bm{r}} \times \bm{\nabla} c_B(\bm{r})\Big], \label{eq:T}
\end{align}
where $\mathbf{I}$ is the unit dyadic,
$\hat{\bm{r}}$ is the outward normal unit vector from the sphere surface,
$c_B(\bm{r})$ is the concentration of species $B$ at $\bm{r}$,
$b$ is the slip length,
$\zeta_t^\circ$ and $\zeta_r^\circ$ are the translational and rotational friction coefficients for perfect stick boundary condition ($b=0$),
$H_h(\theta,\phi) = 1$ if the angles $\theta$ and $\phi$ lie in a surface domain of type $h$, otherwise $H_h(\theta,\phi) = 0$, where $h=C$ or $N$, and $\lambda_{h}= k_{B}T(\lambda_h^{(1)} + b \lambda_h^{(0)})/\eta$ with
\begin{equation}
\lambda_{h}^{(n)} = \int_0^{\infty} dr\:r^{(n)} [{\rm e}^{-U_{h B}(r)/k_{\rm B}T} - {\rm e}^{-U_{h A}(r)/k_{\rm B}T}  ],
\end{equation}
where $U_{hk}(r)$ is the interaction potential between fluid particles of species $k=A,B$ and motor surface of type $h$. While perfect stick boundary conditions are usually considered for micrometric and larger particles, partial slip boundary conditions apply for colloids with hydrophobic interactions and have been suggested to give rise to enhancement of interfacially driven transport phenomena~\cite{ajdari:06}. The colloidal particles considered in the simulations presented below have repulsive interactions with the fluid and satisfy partial slip boundary conditions.

The translational $\zeta_{t}$ and rotational $\zeta_{r}$ friction coefficients of the spherical colloid are
\begin{equation}\label{eq:zeta}
\zeta_{t} = \zeta_t^\circ \frac{1+2b/R_M}{1+3b/R_M}, \qquad
\zeta_{r} = \frac{\zeta_r^\circ}{1+3b/R_M},
\end{equation}
and it then follows that the diffusiophoretic translational and angular velocities are given by
\begin{equation}\label{eq:velocities}
\bm{V} = \bm{F}/\zeta_{t},  \qquad
\bm{\Omega} = \bm{T}/\zeta_{r}.
\end{equation}

In this paper we consider motor dynamics for two situations: (1) the interactions of the $A$ and $B$ particles with motor are the same for the $C$ and $N$ domains, and (2) these interactions differ for the two domains.
For the former case we have $\lambda_C = \lambda_N = \lambda$ and the formulas for the linear and angular velocities reduce to
\begin{align}
  \bm{V} = \frac{\lambda}{4\pi R_M^2(1+2b/R_M)} \:\int_S dS  \:\Big[ (\mathbf{I} - \hat{\bm{r}}\hat{\bm{r}}) \cdot {\bm{\nabla}} c_B(\mathbf{r})\Big],
\end{align}
\begin{align}
  \bm{\Omega} = \frac{3\lambda}{8\pi R_M^3}\:\int_S dS \:  \Big[ \hat{\bm{r}} \times \bm{\nabla} c_B(\mathbf{r})\Big].
\label{eq:Omegad_uni}
\end{align}
For the latter case, considering only the $C$ domain to be responsible for self propulsion, one has $\lambda_N = 0$ and the expressions for the velocities become
\begin{align}
 \bm{V} = \frac{\lambda_C}{4\pi R_M^2(1+2b/R_M)} \:\int_S dS \:H_C(\theta,\phi)  \nonumber \\
 \times \Big[ (\mathbf{I} - \hat{\bm{r}}\hat{\bm{r}}) \cdot {\bm{\nabla}} c_B(\mathbf{r})\Big],\label{eq:Vd}
\end{align}
\begin{align}
 \bm{\Omega} = \frac{3\lambda_C}{8\pi R_M^3}\:\int_S dS \:H_C(\theta,\phi) \: \Big[ \hat{\bm{r}} \times \bm{\nabla} c_B(\mathbf{r})\Big].\label{eq:Omegad}
\end{align}
In this case one can see from Eq.~(\ref{eq:Omegad}) that active rotation takes place
when the distribution of concentration gradients over the $C$ domain must have broken symmetry in the plane of the motor rotation.

\subsection{Concentration field}
The concentration fields that enter the expressions for the linear and angular velocities can be obtained from the solutions of reaction-diffusion equations. For small P{\'e}clet numbers, $\mathrm{Pe}=V R_M/D \ll 1$, where $D$ is the diffusion coefficient of the $A$ and $B$ species, we solve the steady state reaction-diffusion equation for the concentration $c_A$ of species $A$,
\begin{align}
  D\nabla^2 c_A -k_2c_A = 0,
  \label{rxn-diff}
\end{align}
subject to the radiation boundary condition on the motor surface,
\begin{align}
  -\bm{J}\cdot \bm{\hat{r}}\vert_{r=R_c} = \mathcal{R} c_A(r=R_c),
  \label{bc}
\end{align}
where $R_c$ is the radius at which surface chemical reactions occur
and $\bm{J}$ is the flux of species $A$ given by $\bm{J} = -D\nabla c_A$. The concentration $c_B$ of species $B$ can be found from the mass conservation condition, $c_A+c_B=c_0=const$.

In Eq.~(\ref{bc}) $\mathcal{R}$ denotes the chemical reaction rate on the catalytic domain. It may be written as the product of  the rate coefficient $k_0(\theta,\phi)$, which in general depends on angles, and the step function $H_C(\theta,\phi)$,
$\mathcal{R}=k_0(\theta,\phi)H_C(\theta,\phi)/(4\pi R_c^2)$.
Using this form the boundary condition in Eq.~(\ref{bc}) may be written as
\begin{align}
  R_c k_D \frac{\partial c_A}{\partial r} \bigg \vert_{r=R_c}= k_0(\theta,\phi)
  H_C(\theta,\phi) c_A (r=R_c),
  \label{bc2}
\end{align}
where the Smoluchowski diffusion-controlled rate coefficient is given by $k_D=4\pi R_c D$.

\begin{figure}[t!]
  \centering
  \includegraphics[scale=0.9,angle=0]{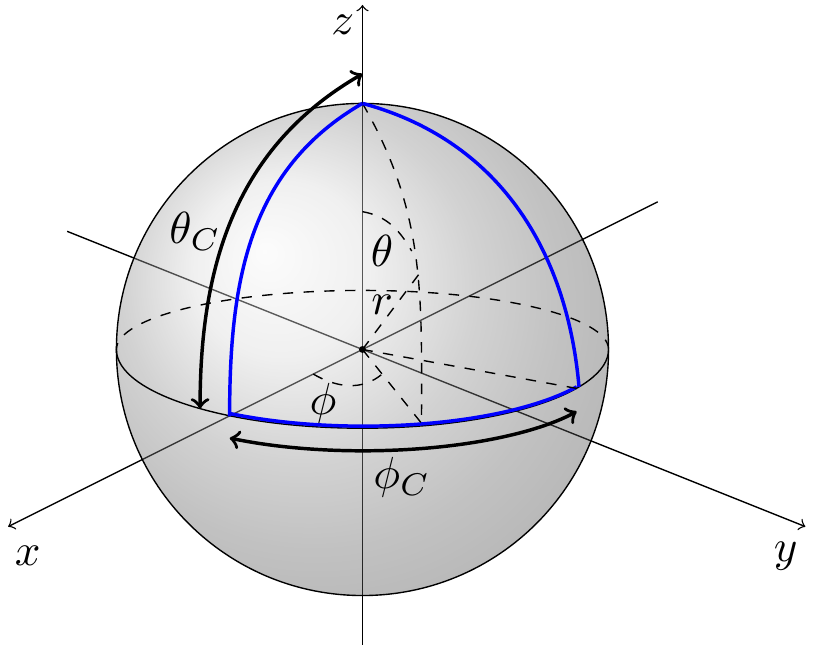}
  \caption{A model of the motor with a catalytic domain specified in spherical polar coordinates ($r$,
    $\theta$, $\phi$) with polar angle $\theta_C$ and azimuthal angle
    $\phi_C$ with respect to the Cartesian coordinates ($x$, $y$, $z$).
    The small region surrounded by the blue lines ($0 <\theta< \theta_C$ and
    $0 < \phi < \phi_C$) is the catalytic domain while the rest of the surface is noncatalytic.
    }
  \label{fig:model}
\end{figure}

The reaction-diffusion equation (\ref{rxn-diff}) in spherical polar coordinates, ($r$,$\theta$,$\phi$), can be solved
by separation of variables and the solution for $c_B$ can be written as
\begin{align}
  c_B = c_0\sum_{n=0}^{\infty}\sum_{m=-n}^{n}A_{nm}\frac{R_n(r)}{R_n(R_c)}Y_{nm}(\theta,\phi),
  \label{conc}
\end{align}
with radial part
\begin{align}
  R_n(r) = \sqrt{\frac{\pi}{2\kappa_2 r}}K_{n+\frac{1}{2}}(\kappa_2 r),
\end{align}
where $\kappa_2 = \sqrt{k_2/D}$
and $K_{n+1/2}$ is a modified Bessel function of the second kind.
The angular part is given in terms of spherical harmonics,
\begin{align}
  Y_{nm}(\theta,\phi) = a_{nm}P_{nm}(\theta,\phi)e^{im\phi},
\end{align}
where $P_{nm}$ is an associated Legendre function, and
$a_{nm} = \sqrt{(2n+1)/4\pi}\cdot\sqrt{(n-m)!/(n+m)!}$.
The solution is expressed in a body-fixed frame (see Fig.~\ref{fig:model}).

The unknown coefficients $A_{nm}$ in the concentration field are found by inserting $c_A=c_0-c_B$ determined from Eq.~(\ref{conc}) into the boundary condition~(\ref{bc2}). Use of the orthogonality of the spherical harmonics leads to the following set of equations:
\begin{align}
  \sum_{n=0}^{\infty}\sum_{m=-n}^{n} M_{n^\prime m^\prime nm} A_{nm} = \gamma_{n^\prime m^\prime},
  \label{mat1}
\end{align}
where
\begin{align}
  M_{n^\prime m^\prime nm} = \alpha_{n^\prime m^\prime nm} + \beta_{n^\prime m^\prime nm}, \nonumber
\end{align}
\begin{align}
  \alpha_{n^\prime m^\prime nm} = \Big\{\kappa_2 R_M\frac{R_{n+1}(R_c)}{R_n(R_c)} - n \Big\}\delta_{n^\prime n}\delta_{m^\prime m}, \nonumber
\end{align}
\begin{align}
  \beta_{n^\prime m^\prime nm} = \frac{1}{k_D} \int_{0}^{2\pi}\int_{0}^{\pi}
  k_0(\theta,\phi) H(\theta,\phi) Y_{nm}Y_{n^\prime m^\prime}^*
  \sin\theta d\theta d\phi,\nonumber
\end{align}
\begin{align}
  \gamma_{n^\prime m^\prime} = \frac{1}{k_D} \int_{0}^{2\pi}\int_{0}^{\pi}
  k_0(\theta,\phi) H(\theta,\phi) Y_{n^\prime m^\prime}^*
  \sin\theta d\theta d\phi,
  \label{matco}
\end{align}
where $\delta_{nm}$ is the Kronecker delta. If $H_C(\theta,\phi)$ is expanded in series of spherical
harmonics, one can obtain explicit expressions for the coefficients
$A_{nm}$ in terms of the coefficients in this series expansion~\cite{lisicki:18}.

For small $\kappa_2$, the radial part of Eq.~(\ref{conc}) reduces to
\begin{align}
  R_n(r) = \frac{\pi(2n)!}{n!}\Big(\frac{1}{2\kappa_2 r}\Big)^{n+1}
\end{align}
and in the limit of vanishing fluid phase reactions ($k_2\rightarrow 0$), corresponding to a source for fuel particles and a sink for product particles far from the colloid, the concentration field of $B$ particles becomes
\begin{align}
  c_B = c_0\sum_{n=0}^{\infty}\sum_{m=-n}^{n}A_{nm}\Big(\frac{R_c}{r}\Big)^{n+1}Y_{nm}(\theta,\phi),
\end{align}
where the coefficients $A_{nm}$ are determined by Eqs.~(\ref{mat1}) and (\ref{matco}) except that
$\alpha_{n^\prime m^\prime nm}$ is given by
\begin{align}
  \alpha_{n^\prime m^\prime nm} = (n+1) \delta_{n^\prime n}\delta_{m^\prime m}.
\end{align}

If the catalytic domain has form of a spherical triangle, it can be specified by the angles $\theta_C$ and $\phi_C$
as the region where $0 \le \theta \le \theta_C$ and $0 \le \phi \le \phi_C$
(see Fig.~\ref{fig:model}).
Since $H_C(\theta,\phi) = 1$ for ($0 \le \theta \le \theta_C$, $0 \le \phi \le \phi_C$)
and 0 otherwise,
$\beta_{n^\prime m^\prime nm}$ and $\gamma_{n^\prime m^\prime}$ in Eq.~(\ref{matco}) can be written as
\begin{align}
  &\beta_{n^\prime m^\prime nm} = \frac{1}{k_D}
  \int_{0}^{\theta_C}\int_{0}^{\phi_C} k_0(\theta,\phi) Y_{nm}Y_{n^\prime m^\prime}^*
  \sin\theta d\theta d\phi,\nonumber\\
  &\gamma_{n^\prime m^\prime} = \frac{1}{k_D}
  \int_{0}^{\theta_C}\int_{0}^{\phi_C} k_0(\theta,\phi) Y_{n^\prime m^\prime}^*
  \sin\theta d\theta d\phi.
  \label{eq:matco_sph}
\end{align}
For a Janus motor with a hemispherical catalytic cap, one has $\theta_C=\pi/2$ and $\phi_C=2\pi$.
If the catalytic activity is uniform, i.e. $k_0(\theta,\phi)=\tilde{k}_0=const$,
one sees  that only the $m=m^\prime=0$ term remains in Eqs.~(\ref{mat1}),~(\ref{matco}) and~(\ref{eq:matco_sph})
since $M_{n^\prime m^\prime nm}=0$ for $m \neq m^\prime$ and
$\gamma_{n^\prime m^\prime}=0$ for $m^\prime \neq 0$, which gives $A_{nm}=0$ for all $n$ and $m \ne 0$.
Hence $\alpha_{n^\prime m^\prime nm}$, $\beta_{n^\prime m^\prime nm}$,
and $\gamma_{n^\prime m^\prime}$ for $m=m^\prime=0$ are given by
\begin{align}
  &\alpha_{n^\prime 0 n 0} = a_{n0}\Big\{\kappa_2 R_c\frac{R_{n+1}(R_c)}{R_n(R_c)} - n \Big\}
  \frac{2}{2n+1}\delta_{n^\prime n},\nonumber\\
  &\beta_{n^\prime 0 n 0} = a_{n0}\frac{k_0}{k_D}\int_0^1 P_{n0}P_{n^\prime 0} d\mu,
  \hspace{10pt}
  \gamma_{n^\prime 0} = \frac{k_0}{k_D} \int_0^1 P_{n^\prime 0} d\mu,
\end{align}
where the integrals have analytic expressions~\cite{grad07}.
In this case one obtains the $B$ concentration field as
\begin{align}
  c_B = c_0 \sum_{n=0}^{\infty} a_{n0}A_{n0}\frac{R_n(r)}{R_n(R_c)}P_{n}(\theta,\phi),
\end{align}
which is consistent with previously given solutions~\cite{huang:16,reigh:16janus}.

\subsection{Linear and angular velocities for special cases}\label{sec:special}

\subsubsection{Uniform $\lambda$ ($\lambda = \lambda_C =\lambda_N$)\label{sec:no_angle}}\label{subsec:uniform}

If the interaction potentials have no angular dependence, for instance by setting $U_{CA}=U_{NA} \ne U_{CB}=U_{NB}$ where the $A$ and $B$ particle interaction potentials do not depend on the domain type, one has $\lambda=\lambda_C =\lambda_N$.
The self-propulsion properties and fluid flow fields of Janus motors with hemispherical caps and such interaction potentials were studied previously~\cite{reigh:16janus}.

The propulsion velocities in the body-fixed frame with Cartesian coordinate unit normal vectors
$\{\hat{\bm{x}}, \hat{\bm{y}}, \hat{\bm{z}}\}$ (Fig.~\ref{fig:model}) may be obtained using Eqs.~(\ref{conc}) and (\ref{eq:Vd}). Noting that $m=\pm 1$ is sufficient for this calculation, one finds that the $x$-component of the propulsion velocity, $V_{x} = \bm{V} \cdot \hat{\bm{x}}$, is given by
\begin{align}
  &V_{x} = \frac{\lambda }{4R_M (1+2b/R_M)} \sum_{n=1}^{\infty} c_0 a_{n, 1}(A_{n, 1}-A_{n, -1})
  \nonumber\\
  & \hspace{40pt} \times \int_0^{\pi} \Big(\frac{dP_{n, -1}}{d\theta}\cos\theta
  \sin\theta d\theta + P_{n, 1} \Big) d\theta.
  \label{eq:vx_mid}
\end{align}
The integral in Eq.~(\ref{eq:vx_mid}) may be evaluated to yield $-8\delta_{n,1}/3$ giving
\begin{align}
  V_{x} = \frac{2\lambda}{3R_M (1+2b/R_M)}  c_0 a_{11}(A_{1-1}-A_{11}).
\end{align}
In a similar manner one finds
\begin{align}
  V_{y} &= -\frac{2\lambda i}{3R_M (1+2b/R_M)} c_0 a_{11}(A_{1-1}+A_{11}), \\
 V_{z} &= \frac{2\lambda}{3R_M (1+2b/R_M)} c_0 a_{10}A_{10},
\end{align}
 for the $y$ and $z$ components.

The angular velocity components in the body-fixed frame may be calculated
from Eq.~(\ref{eq:Omegad_uni}) and one finds that they are zero
since the sum of the concentration gradients around the sphere becomes zero.

\subsubsection{\label{sec:angle}Nonuniform $\lambda$ ($\lambda_C > \lambda_N = 0$)}\label{subsec:nonuniform}

Active motor rotation is obtained if interaction potentials depend on type of surface domain. For example, this is the case if $\lambda_C > \lambda_N = 0 $,
for which the linear and angular velocities are given by Eq.~(\ref{eq:Vd}) and Eq.~(\ref{eq:Omegad}). This case encompasses situations where the $A$ and $B$ particles interact differently with the $C$ and $N$ domains ($U_{CB}\ne U_{NB}$, $U_{CA}\ne U_{NA}$).
The components of the propulsion velocity are given by
\begin{align}
  \begin{pmatrix}
    V_{x} \\
    V_{y} \\
    V_{z}
  \end{pmatrix}
  = \sum_{n=0}^{\infty}\sum_{m=-n}^{n}\alpha_{nm}
  \begin{pmatrix}
    E_{nm} \\
    F_{nm} \\
    G_{nm}
  \end{pmatrix}
\end{align}
where $\alpha_{nm} = \lambda_C c_0 a_{nm}A_{nm}/(4\pi R_M(1+2b/R_M))$ and
\begin{align}
  &\begin{pmatrix}
      E_{nm} \\
      F_{nm} \\
      G_{nm}
    \end{pmatrix}
    = \int_{0}^{\theta_C}\frac{d P_{nm}}{d\theta}\sin\theta
    \begin{pmatrix}
      \cos\theta \\
      \cos\theta \\
      \sin\theta
    \end{pmatrix}
    d\theta \nonumber \\
    &\hspace{70pt} \times \int_{0}^{\phi_C}
    \begin{pmatrix}
      \cos\phi \\
      -\sin\phi \\
      -1
    \end{pmatrix}
  \ e^{im\phi}d\phi \nonumber \\
  &\hspace{50pt}
  - im \int_{0}^{\theta_C}  P_{nm} d\theta
  \int_{0}^{\phi_C}
  \begin{pmatrix}
      \sin\phi  \\
      \cos\phi \\
      0
    \end{pmatrix}
  \ e^{im\phi}d\phi.
\end{align}
The components of the angular velocity are
\begin{align}
  \begin{pmatrix}
    \Omega_{x} \\
    \Omega_{y} \\
    \Omega_{z}
  \end{pmatrix}
  = \sum_{n=0}^{\infty}\sum_{m=-n}^{n}\beta_{nm}
  \begin{pmatrix}
    H_{nm} \\
    I_{nm} \\
    J_{nm}
  \end{pmatrix}
\end{align}
where $\beta_{nm} = 3\lambda_C c_0a_{nm}A_{nm}/(8\pi R_M^2)$ and
\begin{align}
  &\begin{pmatrix}
      H_{nm} \\
      I_{nm} \\
      J_{nm}
    \end{pmatrix}
    = -\int_{0}^{\theta_C}\frac{d P_{nm}}{d\theta}\sin\theta 
    d\theta  
    \int_{0}^{\phi_C}
    \begin{pmatrix}
      \sin\phi \\
      \cos\phi \\
      0
    \end{pmatrix}
  \ e^{im\phi}d\phi \nonumber \\
  &\hspace{50pt}
  + im \int_{0}^{\theta_C}  P_{nm}
  \begin{pmatrix}
    \cos\theta \\
    \cos\theta \\
    \sin\theta
  \end{pmatrix}
  d\theta  \nonumber\\
   &\hspace{70pt} \times \int_{0}^{\phi_C}
  \begin{pmatrix}
      -\cos\phi  \\
      \sin\phi \\
      1
    \end{pmatrix}
  \ e^{im\phi}d\phi.
\end{align}

\section{Simulation of motor dynamics}\label{sec:sims}
\begin{figure}[t!]
\centering
\resizebox{1\columnwidth}{!}{%
\includegraphics{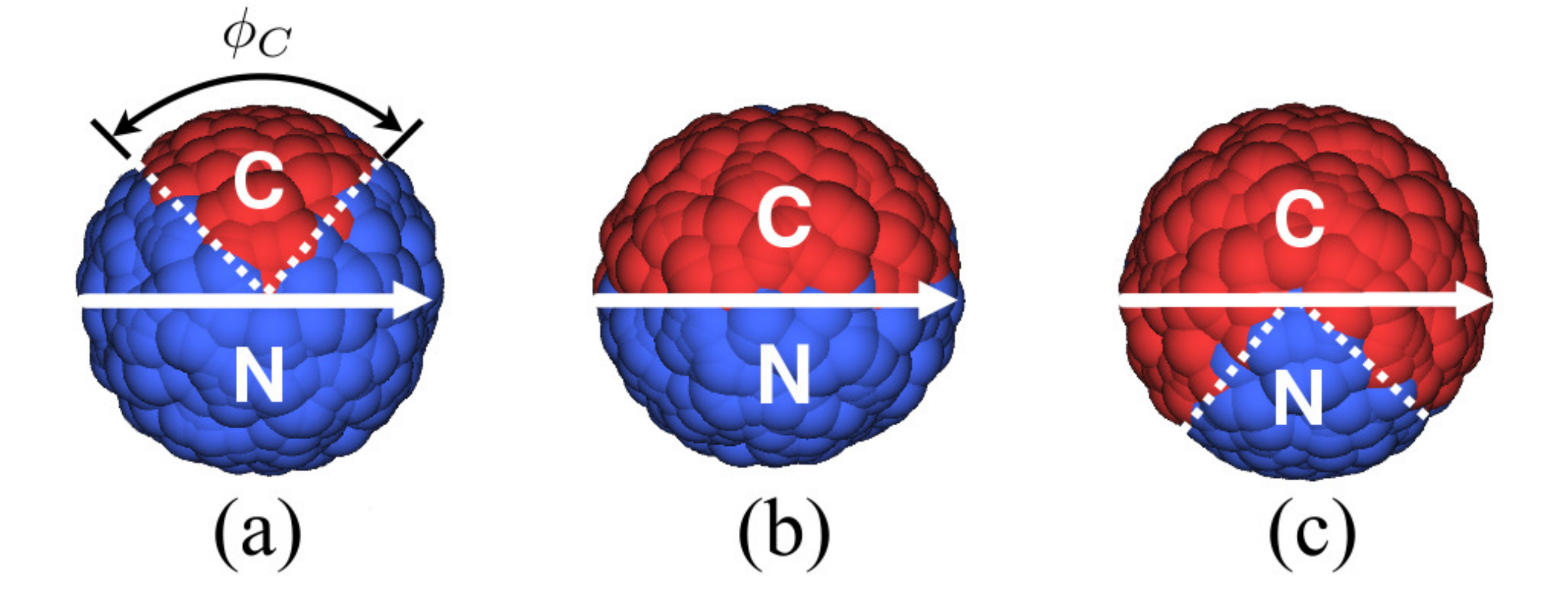}}
\caption{\label{fig:top_view}
  Bead-spring model of motors with catalytic domains:
  Top views from the positive $z$ axis (i.e. the upper hemispheres) in Fig.~\ref{fig:model}.
  The catalytic domain sizes vary as shown in
  (a) $\phi_C = \pi/2$, (b) $\phi_C = \pi$ and (c) $\phi_C = 3\pi/2$,
  with a fixed $\theta_C=\pi/2$.
  When the chemical reaction rate on the catalytic domain depends on
  only the polar angle $\theta$ (for example, $k_0=\tilde{k}_0\cos\theta$,
  where $\tilde{k}_0$ is a constant),
  the motors rotate with the axes indicated by the white solid arrows
  due to the symmetry.
  The catalytic ($C$) and non-catalytic ($N$) parts are indicated by red
  and blue colors respectively and their approximate borderlines are shown in white dotted lines.
}
\end{figure}

Particle-based simulations of motors undergoing active rotational and translational motion were carried out. The Janus motor is made from a collection of small spherical particles (beads) linked to form a large colloidal particle~\cite{debuyl:13,Huang_etal_2018,Huang_etal_2019}. The beads may be chosen to be catalytic or noncatalytic so that Janus colloids with  specifically-shaped catalytic domains can be constructed. In the simulations the Janus motor is placed in a cubic box of linear size $L=50$ with periodic boundary conditions. The fluid is composed of $N_s (= N_A+N_B)$ point particles with $N_A$ fuel ($A$) and $N_B$ product ($B$). The motor is made from $N_b$ beads, residing within a sphere of radius $R_J$ linked by stiff harmonic springs whose the equilibrium lengths are chosen such that the moment of inertia tensor is nearly diagonal (see Fig.~\ref{fig:top_view})~\cite{Huang_etal_2018}. Of the total $N_b$ beads, $N_C$ are catalytic and $N_N$ are noncatalytic, $N_b=N_C+N_N$. The lower hemisphere of the motor is composed solely of $N$ beads, while the upper hemisphere has $C$ and $N$ beads with the size of the $C$ domain determined by the angle $\phi_C$ (see Fig.~\ref{fig:top_view}). The fluid particles evolve by multiparticle collision dynamics~\cite{Malevanets_Kapral_99,Malevanets_Kapral_00,kapral:08,gompper:09} and the coupling between fluid particles and motor beads is described by repulsive Lennard-Jones potentials, $U_{hk}(r) = 4\epsilon_{hk} [(\sigma / r)^{12} - (\sigma/r)^6 + 1/4] \Theta(r-r_c)$, where $h=C,N$, $k=A,B$, $\epsilon_{hk}$ is the interaction energy, $\sigma$ is the effective size of a motor bead and $r_c = 2^{1/6}\sigma$ is the cutoff distance beyond which the potential function vanishes. Chemical reactions, $C+A \to C+B$, may occur when $A$ particles encounter motor $C$ beads, i.e. when $A$ particles pass the cutoff distance $r_c$, and the reaction probability, $p(\bm{r}_i)$, is determined by the position of the nearest catalytic bead at $\bm{r}_i$ in the body frame. The intrinsic reaction rate constant $k_0(\bm{r}_i) = \tilde{k}_0 p(\bm{r}_i) $, where $\tilde{k}_0$ is the intrinsic rate constant for unit reaction probability. To maintain a nonequilibrium steady state, fluid phase reactions $B \stackrel{k_2}{\rightarrow} A$ are carried out using reactive multiparticle collision dynamics~\cite{Rohlf_Faser_Kapral_08}. Further details of the simulation method along with the parameter values used to obtain the results are provided in the Appendix.
All quantities are reported in dimensionless units.

The simulations incorporate thermal fluctuations arising from the particulate nature of the fluid, and deterministic quantities, such as the motor linear and angular velocities discussed in the previous section, are obtained from the simulation data after averaging over an ensemble of realizations of the dynamics.
The linear ($\bm{V}$) and angular ($\bm{\Omega}$) velocities in simulations are initially computed in the laboratory frame.
In the body-fixed frame, the components of the translational (center of mass) velocity, $(V_{x},V_{y},V_{z})$,
are computed from the velocity in the laboratory frame
by $V_{q} = \langle \bm{V} (t) \cdot \hat{\bm{q}} (t)\rangle$
with the principal axes $\hat{\bm{q}}(t) \in \{\hat{\bm{x}}(t),\hat{\bm{y}}(t),\hat{\bm{z}}(t)\}$ (Fig.~\ref{fig:model}),
where brackets denote ensemble averages.
The angular velocity in the body-fixed frame is computed in a similar way.
Letting $\bm{r}_{bi}(t)$ and $\bm{v}_{bi}(t)$ be the position and velocity of the $i$th bead
at time $t$ and $\bm{R}_{cm}(t)$ be the center-of-mass position of the motor in the laboratory frame,
the angular velocity in the laboratory frame is given by
\begin{equation}
  \bm{\Omega}(t) = \sum_{i=1}^{N_b} m_b [\bm{r}_{bi}(t) - \bm{R}_{cm}(t)]
  \times [\bm{v}_{bi}(t) - \bm{V}(t) ]/I,
\end{equation}
where $I=2MR_J^2/5$ is the moment of inertia. Then the components of the ensemble-averaged angular velocity in the moving frame are $\Omega_{q} = \langle \bm{\Omega}(t) \cdot \hat{\bm{q}}(t) \rangle$.

\begin{figure}[t!]
  \centering
  \includegraphics[scale=0.46,angle=0]{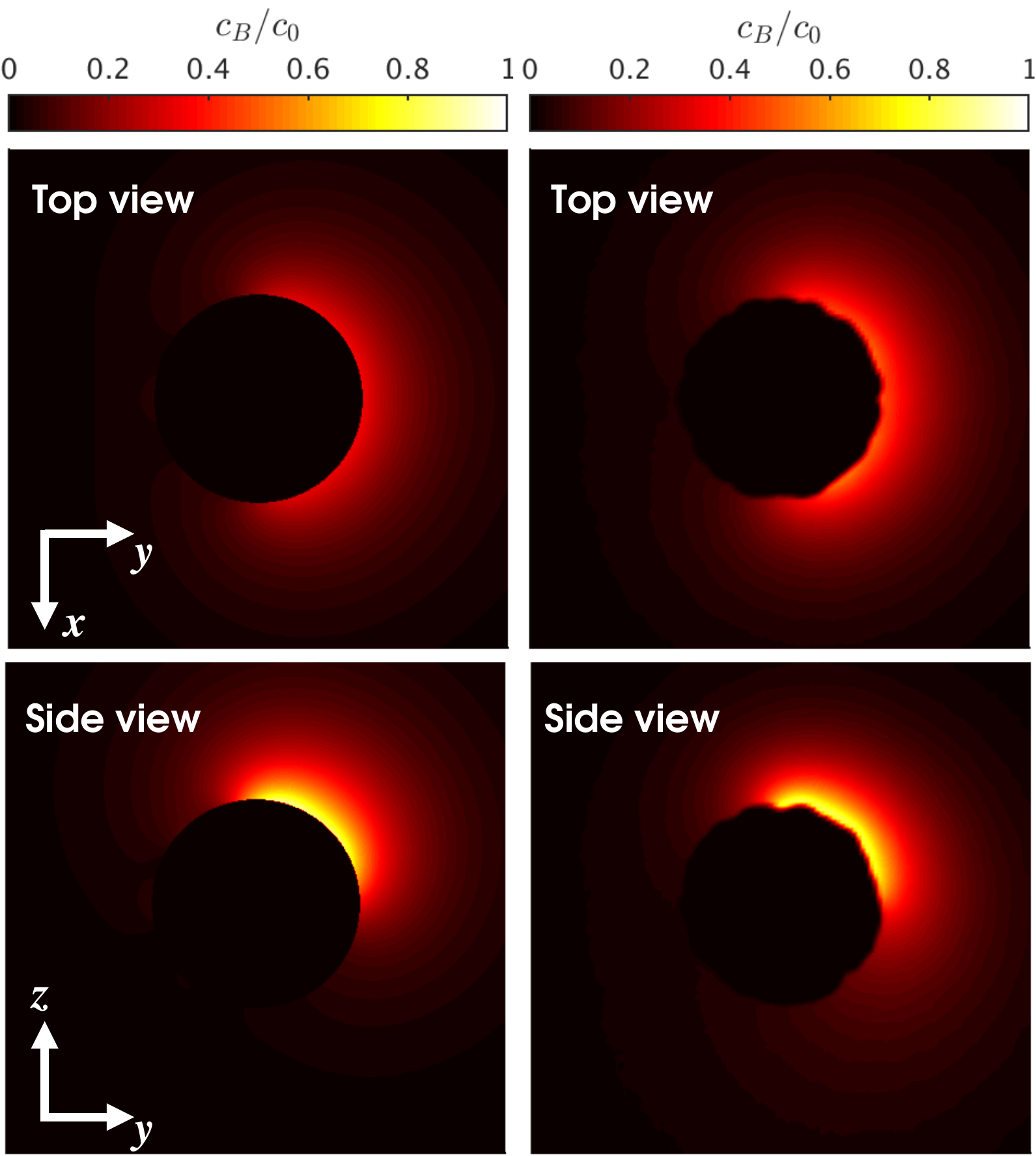}
  \caption{
    Cross sectional views of the normalized chemical $B$ (products) concentration fields
    ($c_B/c_0$) in the vicinity of the motor with a quadrant
    catalytic domain ($\phi_C=\pi$, $\theta_C = \pi/2$, Fig.~\ref{fig:top_view} (b)), where $c_A+c_B=c_0$.
    The left and right columns correspond to the results from
    theory and simulations, respectively.
    The first row is the top view shown in the $xy$ plane
    and the second row is the side view in the $yz$ plane
    (see Fig.~\ref{fig:model}).
  }
  \label{fig:conc}
\end{figure}

Transport properties such as the translational and rotational friction coefficients can be obtained from measurements of their corresponding autocorrelation function expressions. The translational diffusion coefficient $\bar{D}_t$ of an inactive Janus colloid (in the absence of chemical reactions), can be obtained from the time integral of the velocity correlation function,
\begin{equation}\label{eq:Dt}
\bar{D}_t= \frac{1}{3}\int_0^\infty dt \; \langle \bm{V}(t) \cdot \bm{V}(0) \rangle,
\end{equation}
or, equivalently, from the mean square displacement.
(Here the bar notation is used to indicate simulation values for the inactive colloid.)
The translational friction coefficient $\bar{\zeta}_t$ may then be determined from the Einstein relation, $\bar{D}_t=k_BT/\bar{\zeta}_t$. The rotational diffusion coefficient $\bar{D}_r$ can be obtained from the time integral of the orientational correlation function,
\begin{equation}\label{eq:Dr}
\frac{1}{2 \bar{D}_r}= \int_0^\infty dt \; \langle \hat{\bm{u}}(t) \cdot \hat{\bm{u}}(0) \rangle,
\end{equation}
where $\hat{\bm{u}}$ is an orientation vector,
and the rotational friction coefficient $\bar{\zeta}_r$ is given by $\bar{D}_r=k_BT/\bar{\zeta}_r$.

\subsection{Results for active motor translation and rotation}

As discussed in subsect.~\ref{subsec:nonuniform}, the presence of active rotational motion depends on broken symmetry rising from the forms of $\lambda$  and the concentration gradients. In the following, we shall see that these two factors do indeed play important roles in active motor rotational motion.

We consider catalytic domains with shapes shown in Figs.~\ref{fig:model} and~\ref{fig:top_view}.
Specifically the polar angle $\theta_C=\pi/2$ is fixed but the azimuthal angle $\phi_C$ varies from $0$ to $2\pi$, so that the catalytic domain is confined to the upper hemisphere and its size varies with $\phi_C$.
Three choices, $\phi_C = \pi/2$, $\pi$ and $3\pi/2$, are shown in Fig.~\ref{fig:top_view}.

For simplicity, we suppose that the reaction rates depend only on the
polar angle $\theta$; specifically, we take $k_0=\tilde{k}_0\cos\theta$, where $\tilde{k}_0=188.4$.
The energy parameters are chosen to be $\epsilon_{NA}= \epsilon_{NB} = \epsilon_{CA}=1 > \epsilon_{CB}=0.1$,
which gives $\lambda_C= 0.294 \ne \lambda_N=0$ so that $\lambda$ is nonuniform.
For these conditions, the catalytic domain and the concentration field are symmetric
with respect to the bisectional plane $\phi=\phi_C/2$ and rotation is expected to occur about an axis perpendicular to this plane.

\begin{figure}[t!]
  \centering
  \includegraphics[scale=0.5,angle=0]{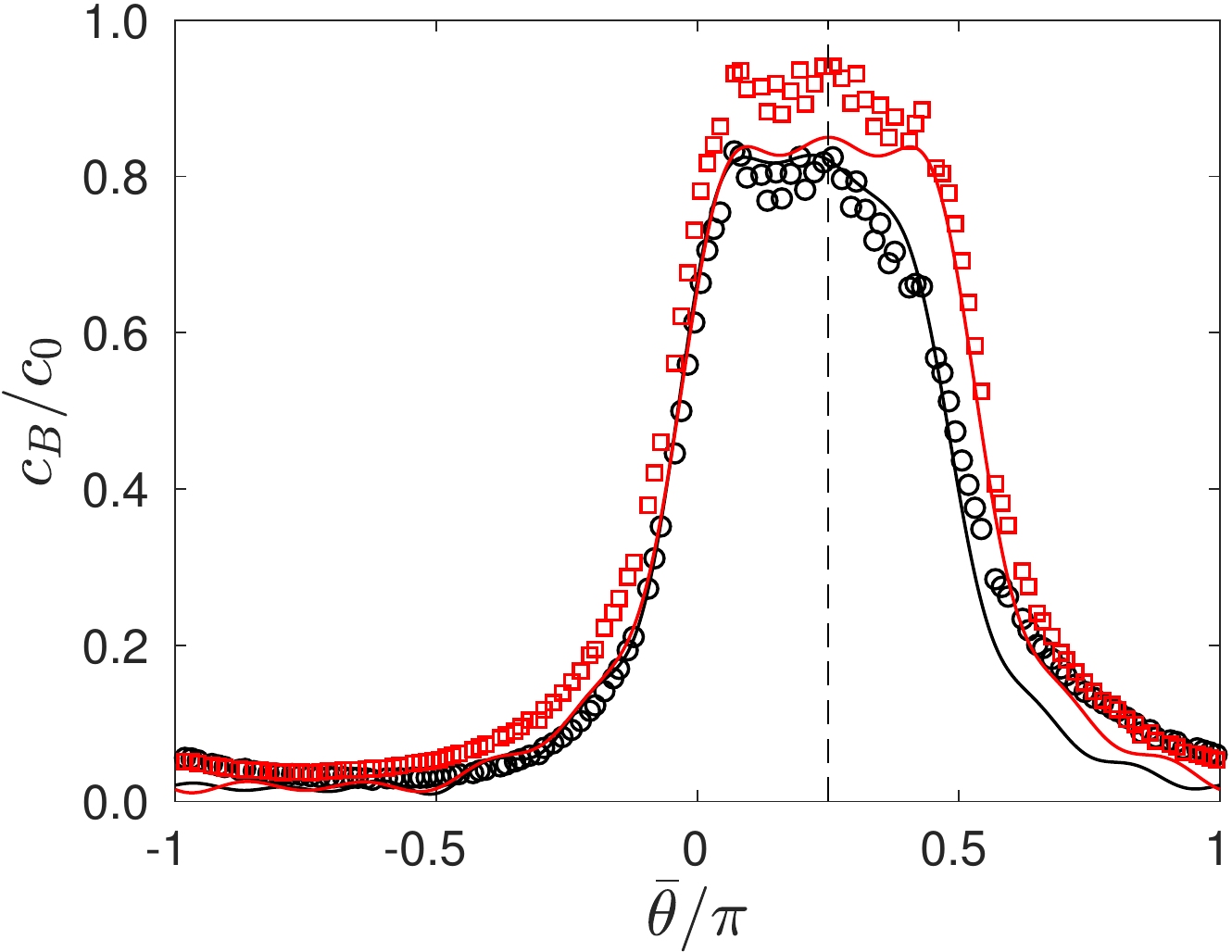}
  \caption{Concentration distributions:
    Quantitative comparison of theory (solid curves) and simulations (symbols) for
    the asymmetric (black circles, $k_0 = \tilde{k}_0 \cos \theta$) and symmetric (red squares, $k_0 = \tilde{k}_0=const$) $B$ concentration field with respect with the angle $\theta = \theta_C/2$ (dashed line).
    The motor has a quadrant catalytic domain as shown in Fig.~\ref{fig:conc}.
    The concentration is plotted along the polar angle in the $yz$ plane
    in the vicinity of the motor surface at $r=r_c=5.12$.
    Here $\bar{\theta} = \theta$ when $\phi=\pi/2$ and
    $\bar{\theta} = -\theta$ when $\phi=3\pi/2$.
    From this, $R_c$ in continuum theory is chosen to be $4.81$
  }
  \label{fig:conc2}
\end{figure}

In addition to presenting simulation results on the active translational and rotational dynamics of the Janus motors, we will compare those with continuum theory. The continuum theory assumes a smooth spherical particle and employs boundary conditions for the fluid velocity and concentration fields on the motor surface that account for the presence of a very  thin boundary layer with length $\delta$ that is much smaller than the particle radius $R_M$ ($\delta << R_M$).
The Janus particle considered in the simulations is not perfectly smooth since it is constructed from beads and the boundary layer within which the fluid particles and colloid interact is of finite size with $\delta/R_M \approx 0.2$.

\begin{figure}[t!]
  \centering
  \resizebox{0.85\columnwidth}{!}{
  \includegraphics[scale=0.52,angle=0]{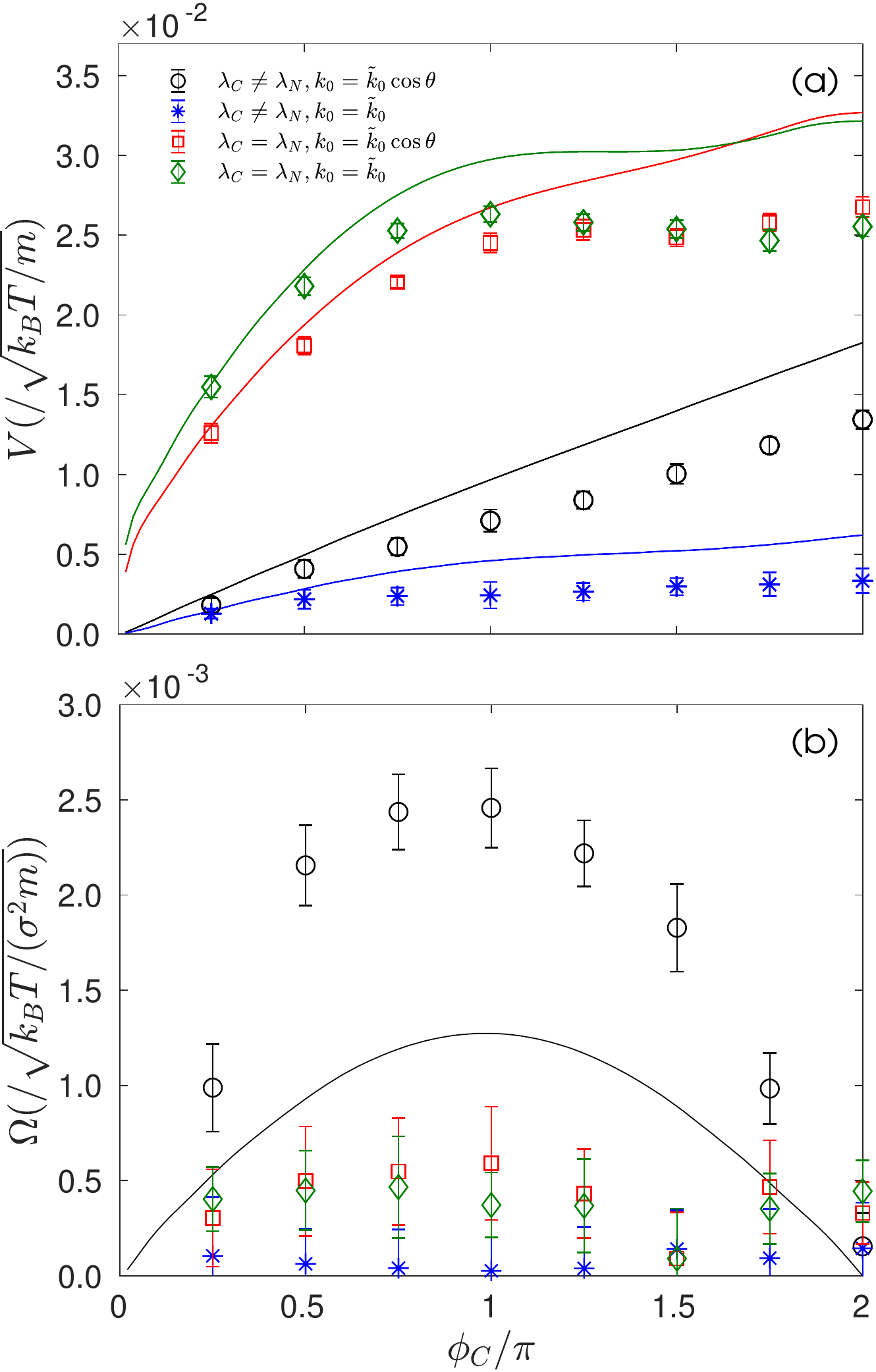}}
  \caption{
    (a) Translational ($V$) and (b) rotational motor velocity ($\Omega$)
    vs the catalytic domain size ($\phi_C$).
    The solid lines and symbol correspond to the theory (Sec.~\ref{sec:special}) and simulations (Sec.~\ref{sec:sims}), respectively.
    Black lines and circles: $\lambda_C \ne \lambda_N$, $k_0=\tilde{k}_0\cos\theta$.
    Blue lines and stars: $\lambda_C \ne \lambda_N$, $k_0=\tilde{k}_0$.
    Red lines and squares: $\lambda_C = \lambda_N$, $k_0=\tilde{k}_0\cos\theta$.
    Green lines and diamonds: $\lambda_C = \lambda_N$, $k_0=\tilde{k}_0$.
  }
  \label{fig:vel}
\end{figure}

\begin{table}[t!]
\centering
\caption{\label{tab:1}
  Simulation results: Ensemble-averaged translational ($\langle V_x\rangle$,$\langle V_y\rangle$,$\langle V_z\rangle$)
  and rotational velocity components ($\langle \Omega_x \rangle$,$\langle \Omega_y \rangle$,$\langle \Omega_z \rangle$)
  of the motors with various catalytic domain sizes ($\phi_C$), as $\theta_C=\pi/2$ is fixed,
  in the body-fixed (moving) frame.
}
\resizebox{1.0\columnwidth}{!}{%
\begin{tabular}{ccccccc}
\hline\hline
$\phi_C$ & $\langle V_x \rangle \times 10^{4}$ & $\langle V_y \rangle \times 10^{4}$ &
$\langle V_x \rangle \times 10^{4}$ & $\langle \Omega_x \rangle \times 10^{4}$ & $\langle \Omega_y \rangle \times 10^{4}$ &
$\langle \Omega_z \rangle \times 10^{4}$   \\
\hline
 $45^{\circ}$ & $\phantom{-}2.81$ &  $1.15$ & $15.6$ & $3.75$ & $-9.14$ & $\phantom{-}1.12$ \\
 $90^{\circ}$ & $\phantom{-}5.77$ &  $7.03$ & $39.0$ & $16.6$ & $-13.7$ & $\phantom{-}0.51$ \\
$135^{\circ}$ & $\phantom{-}3.81$ &  $12.3$ & $52.2$ & $22.3$ & $-9.86$ & $-0.71$ \\
$180^{\circ}$ & $-0.75$ & $15.8$ & $68.8$ & $24.6$ & $-0.31$ & $-0.24$ \\
$225^{\circ}$ & $-0.76$ & $15.5$ & $81.6$ & $21.4$ & $\phantom{-}5.73$ & $-0.96$ \\
$270^{\circ}$ & $-9.17$ & $8.39$ & $99.3$ & $12.4$ & $\phantom{-}13.5$ & $-2.21$ \\
$315^{\circ}$ & $-7.66$ & $2.17$ & $118$ &  $3.66$ & $\phantom{-}9.13$ & $-1.45$ \\
$360^{\circ}$ & $-2.08$ & $0.25$ & $134$ &  $1.54$ & $-0.10$ & $-1.10$ \\
\hline\hline
\end{tabular}
}
\end{table}

Since the interactions between the fluid and colloid are soft repulsive potentials, the effective hydrodynamic radius $R_M$ may differ from the reaction distance $R_c$ and should be determined along with the velocity slip length $b$.
These two quantities $R_M$ and $b$ are estimated from Eq.~\ref{eq:zeta}
using the simulation values of the translational ($\bar{\zeta}_t$) and rotational friction coefficients ($\bar{\zeta}_r$) for the inactive colloids
and assuming $\zeta_t^\circ = 6\pi\eta R_M$ and $\zeta_r^\circ = 8\pi\eta R_M^3$.
The friction coefficients may, in turn, be estimated from measured values of the corresponding diffusion coefficients.
The value of $\zeta_t \sim \bar{\zeta}_t = k_BT/\bar{D}_t \simeq 500$ with $\bar{D}_t \simeq 0.002$ computed in simulations using Eq.~(\ref{eq:Dt}), and the value of the rotational friction coefficient is $\zeta_r \sim \bar{\zeta}_r = k_BT/\bar{D}_r \simeq 3620$ with $\bar{D}_r \simeq 0.000276$ obtained using Eq.~(\ref{eq:Dr}).
Solving the pair of equations~(\ref{eq:zeta}) with the friction coefficients $\zeta_t$ and $\zeta_r$ obtained from simulations, one gets $R_M \simeq 4.61$ and $b \simeq 6.7$.

The reaction radius $R_c$ can be determined from the simulated concentration field. For this purpose we consider a catalytic domain that occupies a quadrant of the spherical surface ($\phi_C = \pi$, $\theta_C = \pi/2$, Fig.~\ref{fig:top_view} (b)). Figure~\ref{fig:conc} shows cross-sectional views in the $xy$ and $yz$ planes
in the body-fixed frame (Fig.~\ref{fig:model}) for the product species in the motor vicinity, while quantitative comparisons are made in Fig.~\ref{fig:conc2}. The asymmetric concentrations (black circles) are plotted as a function of the polar angle in the bisectional symmetry plane $\phi=\phi_C/2$ ($yz$ plane) and at the cutoff distance $r_c=5.12$. Good agreement between simulation and continuum theory is obtained if the reaction radius is taken to be $R_c=4.81$.
This value is close to $R_M \simeq 4.61$.

Figure~\ref{fig:vel} shows the translational and rotational motor velocities
as a function of $\phi_C$. As noted in Sec.~\ref{sec:no_angle},
when the prefactor $\lambda$ is constant ($\lambda_C=\lambda_N$),
the motor has no significant active rotational motion, but only active translation,
whether or not the concentration gradients are asymmetric (red and green colors).
Also, there is no active rotational motion when the concentration gradients are symmetric (blue) as shown in Fig.~\ref{fig:conc2},
although the prefactor $\lambda$ has angle dependence ($\lambda_C \ne \lambda_N$).

When the concentration gradients are asymmetric and $\lambda$ depends on angles,
significant active rotation is observed (black). The theory and simulations are compared for this case and the detailed simulation data are summarized in Table~\ref{tab:1}.
The rotational axes lie in the $xy$ plane so one sees that $\Omega_z$ is small compared to the other components (see Fig.~\ref{fig:model} and Fig.~\ref{fig:conc}).
Since the translational velocity depends on the reaction rate and this scales with the catalytic domain size, $V$ increases with $\phi_C$.
The rotational speed $\Omega$ also increases with the domain size for the same reason
when the domain size is smaller than that for a quadrant-shaped domain ($\phi_C=\pi$).
However, the speed decreases if $\phi_C > \pi$ since the symmetric contributions for rotation cancel.
For example when the domain with $\phi_C=3\pi/2$ (Fig.~\ref{fig:top_view} (c))
is divided to three parts, $0<\phi<\pi/2$, $\pi/2<\phi<\pi$, and $\pi<\phi<3\pi/2$,
the contributions from the first and third parts cancel and only the second part remians, which corresponds to $\phi_C=\pi/2$ (Fig.~\ref{fig:top_view} (a)).
Hence $\Omega$ has a parabolic shape with the maximum at $\Omega = \pi$
and, as expected, the motor does not actively rotate if the entire hemisphere is covered by catalyst ($\phi_C=2\pi$).
It is interesting to note that a motor with a constant $\lambda$ and $k_0$
does not have a significant propulsion velocity increase for $\phi_C>\pi$,
since the length of the borderline of $C$ and $N$ domains which has most
contributions of concentration gradients does not change significantly
(see Fig.~\ref{fig:conc2} and Ref.~\cite{reigh:16janus}).

The continuum theory and simulations are in accord overall for different domain sizes
and various parameter values but the quantitatively differences are more
pronounced for the asymmetric Janus colloids compared to the previous
studies for Janus motors with symmetric catalytic domains~\cite{reigh:16janus,reigh:15dimer,reigh:18dimerform}.

\subsection{Colloid velocity and orientational correlations}
\begin{figure}[t!]
\centering
\resizebox{0.85\columnwidth}{!}{
\includegraphics{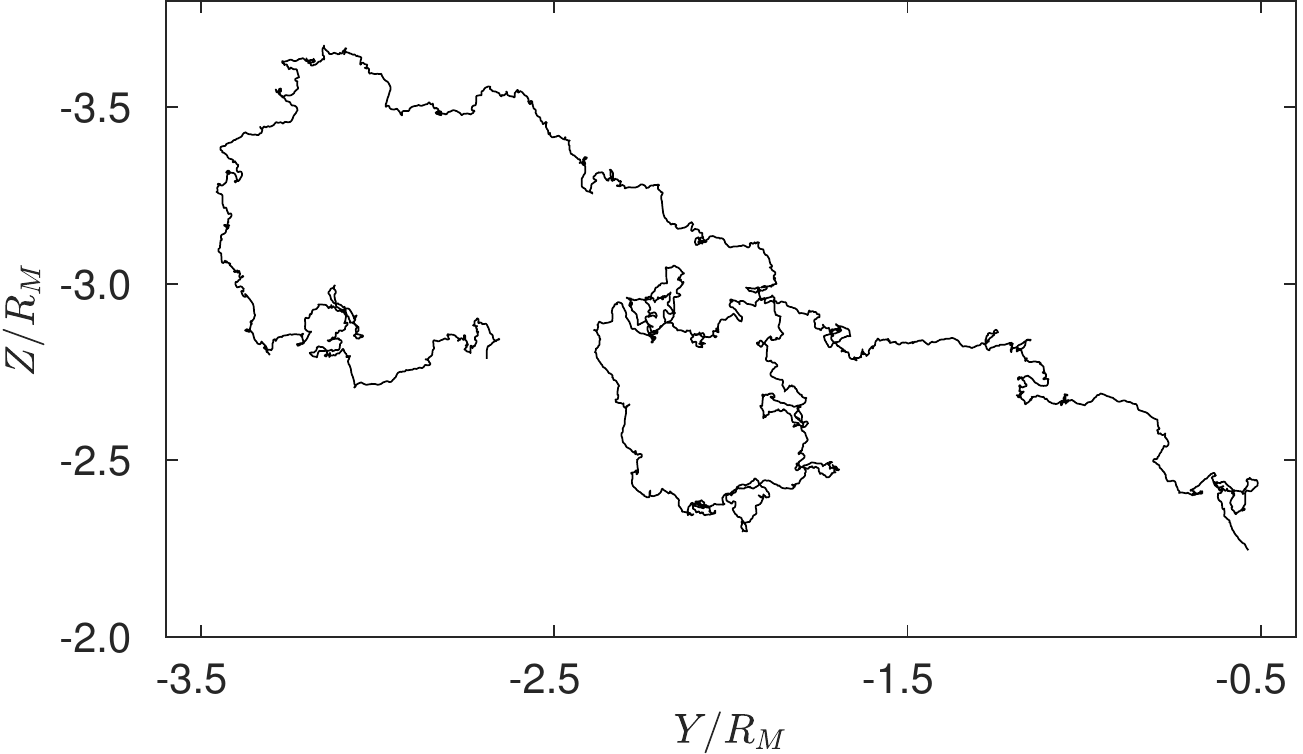}}
\caption{\label{fig:traj}
  A sample trajectory of the motor with a quadrant catalytic domain ($\phi_C =\pi$,
  Fig.~\ref{fig:conc}) projected to the $YZ$ plane in the laboratory frame
  is shown for a time interval
  $t/\tau_r \sim 2.2$.
  The motor rotates counter-clockwise in the $YZ$ plane (Movie S1).
  }
\end{figure}

A sample trajectory from the simulation of the dynamics of a motor with a quadrant catalytic domain ($\phi_C = \pi$) is shown in Fig.~\ref{fig:traj}. One can see that the motor undergoes linear and circular motions as a result of forces and torques induced by diffusiophoretic effects. The velocity and orientational correlation functions provide quantitative information on the active dynamics of the colloid in the presence of thermal fluctuations, and we now consider these quantities. The mean square displacement of the colloid can be expressed in terms of the velocity correlation function,
\begin{align}\label{eq:msd}
\Delta R^2(t)&=\langle (\bm{R}(t)-\bm{R}(0))^2 \rangle \nonumber\\
&=\int_0^tdt_1 \; \int_0^t dt_2\; \langle \bm{V}(t_1) \cdot \bm{V}(t_2) \rangle,
\end{align}
where $\bm{R}(t)$ is the center of mass position of the motor. The effective diffusion coefficient of the active motor can be determined from its long time behavior, $\Delta R^2(t) \sim 6 D_e t$. Equivalently, the time integral of the velocity correlation function, $D_e =\frac{1}{3} \int_0^\infty dt \; \langle \bm{V}(t) \cdot \bm{V}(0) \rangle$, is the analog of Eq.~(\ref{eq:Dt}) for $\bar{D}_t$ for inactive colloids. Likewise the orientational correlations for active rotation are characterized by $\langle \hat{\bm{u}}(t) \cdot \hat{\bm{u}}(0) \rangle$ considered earlier in Eq.~(\ref{eq:Dr}) for inactive colloids.

\begin{figure}[t!]
\centering
\includegraphics[scale=0.55]{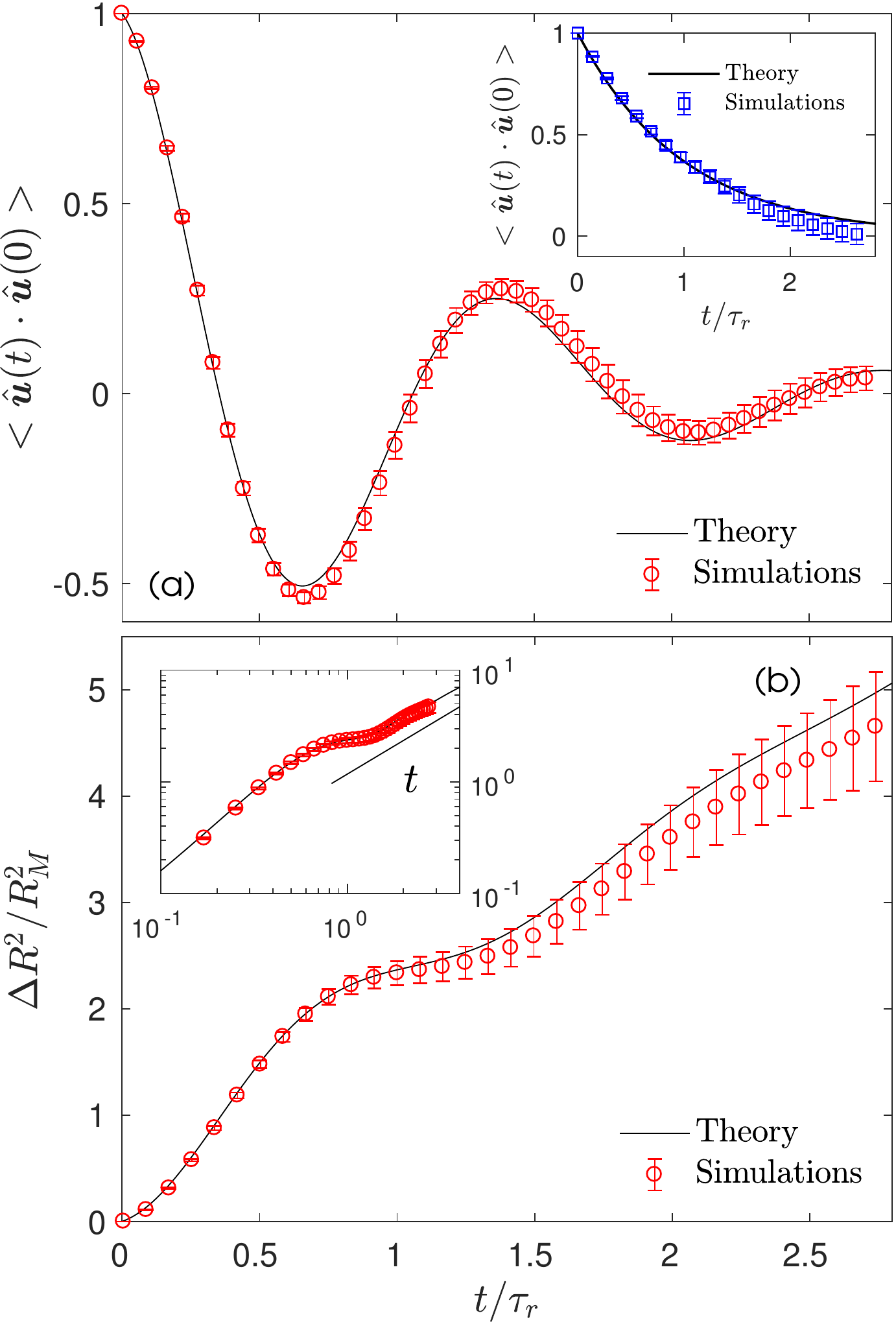}
\caption{\label{fig:ori_corre_msd}
  (a) The orientation correlation function $\langle \hat{\bm{u}}(t) \cdot \hat{\bm{u}}(0) \rangle$
  for a motor with a quadrant-shaped catalytic domain
  ($\phi_C = \pi$, $\lambda_C=0.294 \ne \lambda_N=0$) exhibiting
  translational and rotational motion.
  The inset shows the correlation function for an inactive colloid.
  (b) The mean square displacement $\Delta R^2(t)$ of the motor.
  In the inset the data are plotted in log scales and the labels are the same
  in the main figure. In both of these figures the black solid line and the red circles with error bars correspond to
  the theory and simulations, respectively, except for the inset of (a) that shows inactive motor data in blue.
    }
\end{figure}

Simulations of the orientation correlation function and mean square displacement are shown in Figs.~\ref{fig:ori_corre_msd} (a) and (b), respectively, for colloidal motors with a quadrant-shaped catalytic domain, $\phi_C = \pi$ (Fig.~\ref{fig:conc}),
angle dependent reaction rate, $k_0=\tilde{k}_0\cos\theta$, and interaction energy prefactor, $\lambda_C \ne \lambda_N$.
The orientation correlation function has a decaying oscillatory structure in contrast to the exponential decay for
the inactive motor, as might be anticipated for a colloid with active rotation.

The structures of these results can be understood in the context of overdamped Langevin models for the dynamics~\cite{teeffelen:08}.
In such models, assuming $\zeta_{t,r} \sim \bar{\zeta}_{t,r}$,
the velocity of the motor $\bm{V}(t)$ satisfies
\begin{equation}
  \frac{d}{dt}\bm{R}(t)=\bm{V}(t) = V \hat{\bm{u}}(t) + \bm{V}_f(t),
  \label{langevinV}
\end{equation}
where the fluctuating velocity is a Gaussian white noise process with $\langle  \bm{V}_f(t) \rangle=0$ and fluctuation dissipation relation, $\langle  \bm{V}_f(t_1) \bm{V}_f(t_2)\rangle=2 \bar{D}_t \mathbf{I}\delta(t_1-t_2)$. The orientation of the motor $\hat{\bm{u}}(t)$ obeys
\begin{equation}
  \frac{d}{dt}\hat{\bm{u}}(t)= (\bm{\Omega} +  \bm{\Omega}_f(t)) \times \hat{\bm{u}}(t),
  \label{langevinu}
\end{equation}
where the random angular velocity satisfies $\langle  \bm{\Omega}_f(t) \rangle=0$ and fluctuation dissipation
relation, $\langle  \bm{\Omega}_f(t_1) \bm{\Omega}_f(t_2)\rangle=2 \bar{D}_r \mathbf{I} \delta(t_1-t_2)$.

For the results in Fig.~\ref{fig:ori_corre_msd}, for simplicity in theory~\cite{wittkowski:12} we assume that fluctuations apply to the orientation vector in three dimensions, i.e. $\hat{\bm{u}}(t)=(\cos\vartheta(t),\sin\vartheta(t)\cos\varphi(t),\sin\vartheta(t)\sin\varphi(t))$,
and active rotational motion is restricted in a plane, for example, in the $YZ$ plane in the laboratory frame ($X,Y,Z$), now parallel to the body-fixed frame, while simulations are in full three dimensions.
Then the Langevin equation for the sole angle $\varphi$ is written as
\begin{equation}\label{eq:varphi}
  \frac{d}{dt}\varphi(t) = \Omega + \Omega_{fX}(t),
\end{equation}
where $\Omega_{fX}(t)$ is the $X$ component of the random angular velocity.

Using Eqs.~(\ref{langevinu}) and (\ref{eq:varphi}),
the orientation correlation function neglecting the contributions from $x$ components approximates to
\begin{equation}
  \langle \hat{\bm{u}}(t) \cdot \hat{\bm{u}}(0) \rangle = e^{-t/\tau_r} \cos (\Omega t),
  \label{orien}
\end{equation}
where the rotational relaxation time $\tau_r= 1/(2\bar{D}_r)$~\cite{zwanzig:01,teeffelen:08,ebbens:10}. From
Eqs.~(\ref{eq:msd}), (\ref{langevinV}) and~(\ref{orien}), the mean square displacement takes the form,
\begin{align}
  \Delta R^2 =
  &6 D_{e} t
  - 2 \bigg( \frac{V\tau_r}{1+(\Omega\tau_r)^2} \bigg)^2
  \Big[ (1-(\Omega\tau_r)^2) \nonumber \\
  & -\big\{ (1-(\Omega\tau_r)^2) \cos(\Omega t) - 2\Omega\tau_r \sin(\Omega t)
    \big\}e^{-t/\tau_r}
  \Big],
  \label{msd}
\end{align}
where the effective diffusion coefficient is given
by $D_{e} = \bar{D}_t + V^2\tau_r/\{3(1+(\Omega\tau_r)^2)\}$~\cite{teeffelen:08,ebbens:10}.

Using the diffusiophoretic linear and angular velocities in simulations, $V = 0.007$ and $\Omega =  0.00246$, respectively, the translational diffusion constant $\bar{D}_t \simeq 0.002$ and the orientation relaxation time $\tau_r \simeq 1810$, Fig.~\ref{fig:ori_corre_msd} (a) compares the orientation correlation functions obtained from simulations and theory in the presence and absence of active rotation, whereas panel (b) displays the mean square displacement for the active motors. One finds good agreement.

The mean motor trajectory can be obtained from $\langle \Delta \bm{R}(t) \rangle = \int_0^t \langle \bm{V}(s)\rangle ds$.
By replacing $\hat{\bm{u}}(t)$
by $e^{i\varphi(t)}\sin\vartheta(t)$ approximately in the complex plane 
(where the $YZ$ coordinates are mapped to the complex plane and the $X$ component is neglected)
and using Eq.~(\ref{langevinV}), one obtains
\begin{align}
  \langle {\bm{R}}(t) \rangle \rightarrow \frac{V\tau_r}{\sqrt{1+(\Omega\tau_r)^2}}
  e^{i(\varphi_a+\langle \varphi_0 \rangle)} \big[ 1 - e^{(-1/\tau_r + i\Omega)t} \big],
  \label{eq:trj}
\end{align}
where $\varphi_0 = \varphi(t=0)$ and
$e^{i\varphi_a}=(1+i\Omega\tau_r)/\sqrt{1+(\Omega\tau_r)^2}$.
In the overdamped limit, the trajectory shows a spiral pattern~\cite{teeffelen:08}.
In simulations however the motor experiences fluctuations in three dimensions
and the trajectory projected on the $YZ$ plane is shown in Fig.~\ref{fig:traj}.
The analytical solutions for the translational and rotational motor motion with full three dimensional fluctuations
are more complex~\cite{wittkowski:12}.
If thermal fluctuations are not present then one sees that the motor moves in a circle with a radius $V/\Omega$
in the symmetry ($YZ$) plane by setting $\bar{D}_r=1/(2\tau_r)=0$.

\section{Conclusions}\label{sec:conc}
This theoretical and computational investigation of the active rotational motion of self-diffusiophoretic Janus motors provided quantitative information on the nature of the asymmetrical concentration gradients and solute particle-colloid interaction potentials that give rise to this active motion. Consistent with experimental studies, the asymmetrical concentration gradients can be produced by changing chemical reaction rates locally on the motor surface or using geometric asymmetry~\cite{archer2015,ebbens:10,johnson:17}. Interaction energy asymmetry can arise if the fluid species interact with the catalytic and noncatalytic surface domains through different intermolecular potentials. For given interaction potentials, the domain sizes and shapes, and local variations in the reaction rates, control the motor active angular velocity. For example, in our model, the rotation radius is minimal for a quadrant domain and it increases as the domain size increases or decreases (Fig.~\ref{fig:vel}).

More generally, active orientational motion plays a role in the dynamics of many-motor systems, as well as in scenarios for the control of motor motion for cargo delivery applications. The results provided in this paper should prove useful for such applications involving chemically-powered diffusiophoretic motors. The concepts discussed here for active motor rotation are not restricted to spherical shapes and diffusiophoresis but are applicable to other geometries, such as sphere-dimer motors
and other phoretic mechanisms, although detailed aspects of the description will require modification~\cite{ebbens:10,yang:14,archer2015,johnson:17,robertson:18}.

\section*{Acknowledgements}
S. Y. Reigh greatly thanks S. Dietrich and the Max-Planck-Institute for Intelligent Systems for support.
This work was supported by Brain Pool Program through the National Research Foundation of Korea (NRF) funded by the Ministry of Science and ICT (grant number: NRF-2019H1D3A2A02102052) and
supported in part by the Natural Sciences and Engineering Research Council of Canada and Compute Canada. This project has also received funding from the European Research Council (ERC) under the European Union's Horizon 2020 research and innovation programme  (grant agreement 682754 to EL) and from the DFG within SPP 1726.

\section*{Appendix: Simulation details}\label{appen:sim}
The system is composed of $N_s=N_A+N_B$ fluid $A$ and $B$ particles of mass $m$ and a Janus motor comprising $N_b$ beads of mass $m$ and size $\sigma$ linked by stiff harmonic springs~\cite{Huang_etal_2018}. A chemical reaction, $C+A \to C+B$, may occur when an $A$ particle encounters a motor $C$ bead. The reaction probability $p(\bm{r}_i) = \hat{\bm{z}} \cdot \bm{r}_i/|\bm{r}_i| $ is determined by the position of the nearest $C$ bead at $\bm{r}_i$ in the body frame, where $\hat{\bm{z}}$ is the $z$-axis in the body-fixed frame.

The interactions among fluid particles are accounted for through multiparticle collision dynamics (MPCD). This dynamics consists of streaming and collision steps at discrete time intervals $h_0$ with random shifts of the collision lattice to ensure Galilean invariance~\cite{Malevanets_Kapral_99, Malevanets_Kapral_00, ihle:01}. The system is maintained in a steady state through irreversible reactions $B {\rightarrow} A$ in the fluid solution with reaction rate constant $k_2$. The fluid phase reactions are carried out locally in the collision steps of MPCD.

All quantities are reported in dimensionless units. Length, energy, mass and time are measured in units of $\sigma$, $k_{\rm B}T$, $m$, and $\sigma\sqrt{m/(k_BT)}$, respectively.
The motor beads are homogeneously distributed in a sphere of radius $R_J=4$
with the motor mass $M=m N_b = 2681$ to give neutral buoyancy of the motor.
The moment of inertia $I=2MR_J^2/5=24811$.
The motor reaction radius used in the continuum theory is estimated by comparing the concentration field in simulations at the outer edge of the boundary layer, $r = r_c = 5.12$, which give the reaction radius  $R_c=4.81$ (see Fig.~\ref{fig:conc2}).
The motor is surrounded by a fluid consisting of $N_s=N_A + N_B = 1244219$ particles in the simulation box, excluding the motor volume, which gives a fluid density of $c_0=10$. Multiparticle collisions are carried out in each cell by performing velocity rotations by an angle $\alpha=120^\circ$ about a randomly chosen axis at discrete time intervals $h_0 = 0.1$. The molecular dynamics time step is $\Delta t = 0.01$. The transport properties of the fluid depend on $h_0$, $\alpha$, and $c_0$. The fluid viscosity is given by $\eta = m c_0 \nu = 7.9$, where $\nu$ is the kinematic viscosity. The common diffusion constant of $A$ and $B$ is $D=0.07$. The Schmidt number is $\mathrm{Sc}=\nu/D=13>1$, which ensures that momentum transport dominates over mass transport, the Reynolds number $\mathrm{Re}= c_0V\sigma/\eta < 0.1$ implying that viscosity is dominant over inertia, and the P\'{e}clet number $\mathrm{Pe} = V\sigma/D <1$ indicating diffusion dominates over fluid advection. The energy parameters for interactions between the catalytic beads and $A$ and $B$ species are different ($\epsilon_{CB} = 0.1$, $\epsilon_{CA} = 1$) while they are the same for the noncatalytic beads ($\epsilon_{NA} = \epsilon_{NB} = 1$). We then have, $\lambda_C = k_{\rm B}T(\lambda_C^{(1)} + b \lambda_C^{(0)})/\eta \simeq 0.294$ with $\lambda_C^{(0)}=0.1$ and $\lambda_C^{(1)}=0.476$ and $\lambda_N=0$ since $\lambda_N^{(0)}= \lambda_C^{(1)} = 0$ and the motor moves with the catalytic domain at its head~\cite{Huang_etal_2018}.
The intrinsic reaction rate coefficient $\tilde{k}_0$ is obtained
from comparison of theory and simulations by setting up the rate equations
with radiation boundary conditions, which gives $\tilde{k}_0=188.4$
through averages over multiple realizations.
The reaction rate for the bulk reaction is taken to be $k_2=0.003$.



\providecommand*{\mcitethebibliography}{\thebibliography}
\csname @ifundefined\endcsname{endmcitethebibliography}
{\let\endmcitethebibliography\endthebibliography}{}


\end{document}